\documentclass[aps,prb,twocolumn,showpacs,preprintnumbers,amsmath,amssymb]{revtex4}
\usepackage{epsf}

\begin{document}
\draft
\newcommand{\ve}[1]{\boldsymbol{#1}}

\title{Electron-Polarization Coupling in Superconductor-Ferroelectric Superlattices}
\author{N.~Pavlenko\cite{e-mail} and F.~Schwabl}
\address{Institut f\"{u}r Theoretische Physik T34, Physik-Department
der TU M\"{u}nchen, James-Franck-Strasse, D-85747 Garching
b.~M\"{u}nchen, Germany}
\date{\today}

\begin{abstract}
We present a phenomenological model of periodic ferroelectric-superconductor
(FE-S) heterostructures containing two alternating ferroelectric and
superconducting layers. The interaction at the FE-S contacts is described as a
coupling of the local carrier density of the superconductor with the
spontaneous ferroelectric polarization near the FE-S interface. We obtain a
stable symmetric domain-type phase exhibiting a contact-induced polarization
and the ferroelectric domain structure at temperatures above the bulk
ferroelectric transition temperature. With increasing coupling energy, we find
the appearance of the ferroelectric phase coexisting with the suppressed
superconductivity in the S-film. The system is analyzed for different
thicknesses of the FE- and S-films demonstrating the dramatic change of the
topology of the phase diagrams with a variation of the layers thickness. The
results are expected to shed light on processes occurring in high-temperature
superconducting films grown on perovskite alloy-substrates exhibiting
ferroelectric properties at lower temperatures.
\end{abstract}

\pacs{74.81.-g,74.78.Fk,77.80.-e}

\maketitle

\section{INTRODUCTION}
Epitaxial combination of superconducting films and ferroelectric layers is
promising for construction of effective microwave devices. Especially
advantageous is the application of small-loss high-temperature cuprate
superconductors (YBCO)\cite{vendik1,vendik2} and perovskite ferroelectric
alloys like Ba$_x$Sr$_{1-x}$TiO$_3$ (BST) and Pb(Zr$_x$Ti$_{1-x}$)O$_3$ (PZT),
in heterogeneous miltilayers. As both of these compounds have similar
perovskite-type crystal structure with small lattice mismatch, the well
oriented BST and PZT thin layers can be grown on the surfaces of
YBCO\cite{petrov,ryen,petrov2,keuls}. In addition, the role of strains and
off-stoichiometry structural defects at the BST(PZT)/YBCO-interfaces can be
minimized by the use of growing methods such as laser ablation for
fabrication of the miltilayered structures\cite{petrov}.

Since the 1960s, the strong interest in developing electric field-effect
devices based on ferroelectrics has been stimulated by observations of shifts
of superconducting transition temperature under a switch of the spontaneous
polarization. Superconducting field-effect transistors (SUFETs) cannot be
realized with conventional superconductors since the large coherence lengths
and high carrier density of the latter result in negligibly small effect of
the polarization field \cite{konsin,konsin2}. In contrast, the carrier
density and coherence lengths of high-T$_c$ cuprates (HTS) which are two orders of
magnitude lower than those in conventional superconductors\cite{jenks}, have made
their use in SUFETs very attractive \cite{mannhart1}.

For example, it was demonstrated in Ref.~\onlinecite{larkins,lemanov,lemanov2},
that the critical temperature of YBCO films could  be shifted by about 6~K through
the polarization in the BaTiO$_3$ substrate. As was assumed in
\cite{lemanov,xi,nakajima,mannhart2}, the increase of $T_c^S$ is connected with
the formation of the accumulation layers (with charge carrier density enhanced
due to the polarization directed toward the interface), whereas the
polarization in the opposite direction induces a depletion layer with higher
resistance and lower $T_c^S$. Besides critical temperature, ferroelectric
polarization affects transport properties  of SUFETs leading to a change in
resistance of 9-25~$\%$ for PZT-YBCO and PZT-NdBa$_2$Cu$_3$O$_{7-\delta}$(NBCO)
heterostructures \cite{watanabe,triscone} and
showing polarization-induced memory effects in resistance and current, which
suggests their possible storage applications\cite{watanabe1}.

In the case when the polarization is parallel to the FE-S interface, the use of
the zero-$P$ boundary conditions ($\ve{P}=0$ at the electrode) results in a
weak effect of the FE-layers scarcely affecting the behavior of the S-films. On
the other hand, the influence of the FE-layer should be essentially strong when
the FE-polarization vector $\ve{P}$ is perpendicular to the FE-S interface. Due
to the specific perovskite-like structure, the properties of the high-T$_c$ materials
are believed to be very close to the perovskite-type ferroelectrics. This fact
stimulated discussions about a coexistence of ferroelectricity and
superconductivity and a possible ferroelectricity of
HTS-compounds\cite{muller,bussmann,vezzoli}. As a consequence, the
ferroelectric polarization has been suggested to be nonzero at the interface with
the HTS-layer with even the possibility of penetration inside the
superconductor\cite{vendik4}.

Recent studies of ferroelectric-HTS layered structures were focused on
investigations of the role of free $P$-boundary conditions (non-zero $P$) on
dielectric properties of FE-layers\cite{vendik1,vendik2,vendik4}. However,
little attention has been paid to the behavior of superconducting films as
well as to the effect of the FE-S interface on ferroelectric structural
transformations.

In this work, we develop a phenomenological model based on the Ginzburg-Landau
theory, for the FE-layers sandwiched between superconducting films. As
compared to the microscopic approach proposed in Ref.~\onlinecite{pavlenko}
where the S-layer was described as two S-planes with a possibility of a charge
transfer between the planes and the interior of the layer, the proposed model
allows one to study directly the behavior of the system for different thicknesses
of the S-layer in the vicinity of the ferroelectric transition temperature in the
FE-film, which is assumed to be close to the transition temperature of the
superconductor. We note that for HTS-compounds the condition $\xi_S(T) \gg \xi_S^0$
for the Ginzburg-Landau theory is valid over a much wider temperature range than in
conventional superconductors \cite{tinkham}. We focus primarily on the case when
the penetration depth of the polarization-induced electric field is larger
than the superconducting coherence length ($l_{TF}/\xi_S^0 \gg 1$), which is also
typically realized in HTS layers.

The paper is organized as follows. In section~\ref{model}, we introduce the
Ginzburg-Landau energy functional for the FE-S heterostructure. We derive the
Ginzburg-Landau equations and elucidate the nature of the boundary conditions.
In section~\ref{phases}, we analyze in details the results for the case when
the bulk ferroelectric transition temperature lies below the temperature of the
superconducting transition. We discuss possible phases which can be stabilized
in the system depending on the values of the temperature and the coupling
energy. Corresponding phase diagrams are analyzed in section~\ref{diagrams} for
different thicknesses of the FE- and S-layers. We discuss here also the
thicknesses-dependences of the ferroelectric transition temperature. Concluding
remarks are presented in section~\ref{conclusions}.

\section{THE MODEL \label{model}}

\begin{figure}[htbp]
\epsfxsize=8.5cm \centerline{\epsffile{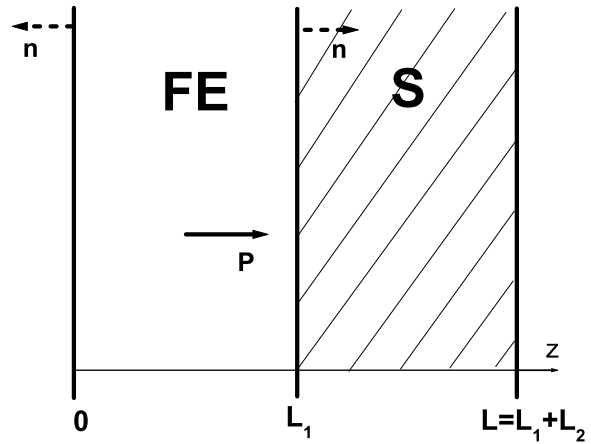}} \caption{\label{fig0}Scheme
of the periodic FE-S multilayer structure.}
\end{figure}

We consider a system containing periodically alternating ferroelectric (FE)
layer sandwiched between superconducting (S) films (shown in Fig.~\ref{fig0})
infinitely extended in the $x$ and $y$ directions. We assume here that
the polarization $\ve{P}=(0,0,P(z))$ is
directed perpendicular to the FE-S interface. The thicknesses of the FE-
and S-layers are given by $L_1$ and $L_2$ respectively, and thus the total width
of the supercell is: $L=L_1+L_2$. Generally, the whole sandwich
structure can be described by the following Ginzburg-Landau functional
\begin{equation}\label{g_l}
  F=F_{FE}+F_S+F_{int}.
\end{equation}
The ferroelectric part of the free energy (\ref{g_l}) is given by
\begin{equation}\label{f_ferr}
  F_{FE}=\int_0^{L_1} dz \left[ \frac{1}{2}a_F P^2(z)+ \frac{1}{4}b_F P^4(z)+
  \frac{1}{2}c_F \left(\frac{dP(z)}{dz}\right)^2 \right],
\end{equation}
where $a_F=a_F^0(T-T_c^F)$ is the inverse of the ferroelectric susceptibility
and $T_c^F$ denotes the bulk Curie temperature. For simplicity, we do not include
the sixth-order
term into the $P$-expansion (\ref{f_ferr}), restricting our
analysis to second-order phase transformations. Note that since the
polarization depends on $z$ only, the last term in (\ref{f_ferr}) simplifies to
$\nabla \ve{P}=\frac{dP(z)}{dz}$.

For the superconducting S-film without magnetic field the free energy can be written as
\begin{equation}\label{f_s}
 F_{S}=\int_{L_1}^{L} dz \left[ \frac{1}{2}a_S |\psi(z)|^2+ \frac{1}{4}b_S |\psi(z)|^4+
  \frac{\hbar^2}{2m e} \left|\frac{d\psi(z)}{dz}\right|^2 \right],
\end{equation}
where $a_S=a_S^0(T-T_c^S)$ with $T_c^S$ denoting the bulk superconducting
transition temperature, and $m$, $e$ and $|\psi(z)|^2$ are the mass, charge and
the local density of the Cooper pairs respectively. Similarly to (\ref{f_ferr})
we assume in (\ref{f_s}) that $\psi(\ve{r})$ depends on $z$ only, i.e.,
$\psi=\psi(z)$. Note also that without a magnetic field the phase $\phi=const$,
and the vector potential $A=\frac{\hbar c}{2e}\frac{d\phi}{dz}=0$ (see the
Appendix). The Ginzburg-Landau equations in this static case contain only the
terms with $|\psi(z)|$ and below for simplicity we use the notations
$|\psi|=\psi$.

The interface part of the free energy (\ref{g_l}) describes the interaction
between the spontaneous polarization and the superconducting charge near the
interface. In order to derive $F_{int}$, we consider first the interaction of
$P$ with an electric field $\ve{E}=(0,0,E(z))$ associated with the polarization
distribution \cite{landau}
\begin{eqnarray}\label{p_e_d}
&& \int_0^{L_1}dz \left (-\frac{1}{2}\ve{P}\ve{E}+\frac{D(z) E(z)}{4\pi}\right)\nonumber\\
&&=\frac{1}{2}\int_0^{L_1} dz P(z) E(z)+\frac{1}{4\pi}\int_0^{L_1} E^2(z),
\end{eqnarray}
where $\ve{D}=\ve{E}+4\pi\ve{P}=\varepsilon \ve{E}$ is the electric
displacement. The field $\ve{E}$ can be determined from the Maxwell equation
inside the FE-layer $div \ve{D}=div(\ve{E}+4\pi \ve{P})=0$. The integration in
the vicinity of the contacts gives us near $z=0$
\begin{eqnarray}\label{Ed_0}
E(z)-E(0^+)=-4\pi (P(z)-P(0^+)),
\end{eqnarray}
and near $z=L_1$
\begin{eqnarray}\label{Ed_L1}
E(L_1^-)-E(z)=-4\pi (P(L_1^-)-P(z)).
\end{eqnarray}
In contrast to the normal metal electrodes where $\ve{P}=0$ and $\ve{E}=0$, we
employ the following two concepts for the FE-S interface: (i) ${P}(0^-),
P(L_1^+) \ne 0$ (see the corresponding discussion in Introduction); (ii) an
electric field can penetrate inside the superconductor ($E(z\ge 0^-), E(z\ge
L_1^+) \ne 0$) and thus substantially influence the superconducting
properties\cite{lemanov,jenks,artemenko}. Using these assumptions together with
the boundary conditions for $\ve{D}$ \cite{jackson}
\begin{equation}\label{D_cont}
  (\ve{D}|_S-\ve{D}|_{FE})\cdot \ve{n}=4\pi\sigma,
\end{equation}
where $\sigma$ denotes the charge density induced by the polarization at the
FE-S contact and $\ve{n}$ is the unit vector directed perpendicular to the
surface of the ferroelectric into the S-layers, we obtain from (\ref{Ed_0}) and
(\ref{Ed_L1})
\begin{eqnarray}\label{Ed_b}
&&E(0^-)+4\pi P(0^-)=-4\pi \sigma(0^-),\\
&&E(L_1^+)+4\pi P(L_1^+)=4\pi \sigma(L_1^+), \nonumber\\
&&E(z)=-4\pi P(z), \quad 0^+ \le z \le L_1^-.\nonumber
\end{eqnarray}

To derive the expressions for $E(z)$ penetrating inside the S-film at a
distance of the Thomas-Fermi charge screening length $l_{TF}$, we consider the
Poisson's equation near the right ($L_1 < z < L_1+l_{TF}$) and left ($-l_{TF} <
z < 0$) FE-S contacts \cite{jackson}
\begin{equation}\label{poisson}
 l_{TF}^2 \frac{d^2\varphi}{dz^2}=\varphi,
\end{equation}
where $-\nabla_z \varphi(z)=E(z)$, with the boundary conditions for the
electrostatic potential $\varphi$
\begin{eqnarray}\label{pois_bound}
-\nabla_z \varphi(L_1^+)=E(L_1^+), && -\nabla_z \varphi(0^-)=E(0^-).
\end{eqnarray}
Here $l_{TF}^2=\frac{\varepsilon_F}{6\pi n_0 e^2}$ is larger for systems with
the lower mobile charge carrier concentration $n_0$, and $\varepsilon_F$ is the
local Fermi energy. The solutions obtained from (\ref{poisson}),
(\ref{pois_bound}) decay exponentially inside the right S-film ($z>L_1$)
\begin{eqnarray}\label{E_pen_r}
\varphi(z)=E(L_1^+)l_{TF} \cdot {\rm e}^{\frac{L_1-z}{l_{TF}}}, &&
E(z)=E(L_1^+) {\rm e}^{\frac{L_1-z}{l_{TF}}}
\end{eqnarray}
and the left S-film ($z<0$)
\begin{eqnarray}\label{E_pen_l}
\varphi(z)=-E(0^-)l_{TF}\cdot {\rm e}^{\frac{z}{l_{TF}}}, && E(z)=E(0^-) {\rm
e}^{\frac{z}{l_{TF}}}.
\end{eqnarray}

The electric field penetration depth $l_{TF}$ in HTS compounds is shown to be
in the range of about $0.5$~nm (Ref.~\onlinecite{mannhart1}), which is
significantly larger than $l_{TF}\sim 0.02$~nm in the normal metal electrodes.
Note that the total charge in the S-layer: $Q=\int_{L_1}^L dz \rho(z)=0$
includes the electron and background contributions:
$\rho(z)=\rho_b(z)+\rho_e(z)$ with the electron charge consisting of the normal
and superconducting components: $\rho_e(z)=\rho_e^n(z)+\rho_e^s(z)$ according
to the two-fluid model of superconductivity \cite{tinkham}. Since we study in
this work the superconducting contribution $\rho_e^s(z)$, we assume here that
the induced surface density $\sigma$ is described by a local increase/decrease
of the superconducting charge density
\begin{eqnarray}\label{sigma}
&&\sigma(L_1^+)=\chi_0 \int_{L_1^+}^{L_1 +l_{TF}} |\psi(z)|^2 dz
=\gamma_0 |\psi(L_1^+)|^2,\nonumber\\
&&\sigma(0^-)=\gamma_0 |\psi(L^-)|^2 ,
\end{eqnarray}
where $\gamma_0 \sim \chi_0 l_{TF}$ is the coupling coefficient characterizing
the change of the charge carrier density near the contacts due to the
spontaneous polarization in the FE-layer.

From the expressions (\ref{Ed_b}), (\ref{sigma}), and (\ref{p_e_d}) we obtain
$F_{int}=F_{int}^0+F_{FE-S}$, where the interaction energy inside the FE-layer
\begin{eqnarray}\label{f_int}
&&F_{int}^0= 2\pi\int_0^{L_1} dz P^2(z).
\end{eqnarray}
The electrostatic interaction of the penetrating field $E(z)$ given by
(\ref{E_pen_r})-(\ref{E_pen_l}), with the superconducting charge $\rho_e^s$
near the contacts ($L_1 < z < L_1+l_{TF}$) and ($-l_{TF} < z < 0$) is
determined by
\begin{eqnarray}\label{f_fe_s0}
F_{FE-S}=\left(\int_{L_1}^{L_1+l_{TF}}+\int_{-l_{TF}}^0\right) dz \varphi(z)
|\psi(z)|^2.
\end{eqnarray}
To calculate the integrals in (\ref{f_fe_s0}), we use the expansion of
$\psi(z)$ at a small distance $\delta z<l_{TF}\ll L_1$ near the contact $z=L_1$
\begin{equation}
    \psi(z)=\psi(L_1^+)+\frac{d\psi(L_1^+)}{dz} (z-L_1)+\ldots
\end{equation}
and $z=0$
\begin{equation}
    \psi(z)=\psi(0^-)+\frac{d\psi(0^-)}{dz} z+\ldots,
\end{equation}
and the expressions (\ref{E_pen_r}) and (\ref{E_pen_l}) for $\varphi$. As the
result, we get
\begin{eqnarray}\label{f_fe_s}
&&F_{FE-S}= -\frac{\gamma}{2} P(L_1^+)|\psi(L_1^+)|^2+\frac{\gamma}{2} P(0^-)|\psi(L^-)|^2\nonumber\\
&& + \delta (|\psi(L_1^+)|^4+|\psi(L^-)|^4)\nonumber + 2\delta_1
\left(P(L_1^+) |\psi(L_1^+)| \frac{d\psi(L_1^+)}{dz}\nonumber \right.\\
&&\left. -P(0^-)|\psi(0^-)|\frac{d\psi(0^-)}{dz}\right)+\mathcal{O}(\gamma^2),
\end{eqnarray}
where $\gamma/2=4\pi (1-1/{\rm e}) \l_{TF}^2$, $\delta=4\pi(1-1/{\rm
e})\gamma_0 l_{TF}^2 \sim \gamma^{3/2}$ and $\delta_1=4\pi(1-2/{\rm e})l_{TF}^3
\sim \gamma^{3/2}$ . The first term (\ref{f_int}) renormalizes the coefficient
$a_F$ in (\ref{f_ferr}) and leads to a symmetrical reduction of the deviation
of $P$ from its bulk value near the boundaries as was shown in
Ref.~\onlinecite{binder}. As a first step in the analysis of the role of the
interface charge carriers, in this work we focus our attention on the
contribution of the first two interface terms in the $F_{FE-S}$ given by
(\ref{f_fe_s}), neglecting the smaller higher order terms in the
$\gamma$-expansion and disregarding the well studied term (\ref{f_int}). It
should be noted that the contribution (\ref{f_fe_s}) can be derived from a
microscopic model describing an Ising-type FE-layer sandwiched between the
S-planes with BCS-pairing\cite{pavlenko}. In the latter approach, the coupling
energy $-\frac{\partial\varepsilon_l}{\partial R}|_{R=R_0}$ characterizes the
change of the one-electron energy $\varepsilon_l$ in the S-plane due to the
distortion of the nearest atomic group of the FE-layer with the coordinate
$R=R_0+\Delta_R$ (the average distortion $\Delta_R$ becomes nonzero in the
ferroelectric state).

\subsection{The Ginzburg-Landau equations \label{Lan_ginz_eq}}

Minimization of the total free energy $f=\int_{\Omega} d\Omega
(F_{FE}+F_{S}+F_{FE-S})$  where $\Omega$ denotes the ($x$, $y$)-surface, by
variation of the order parameters $P$ and $\psi$ yields the following two
Ginzburg-Landau equations
\begin{eqnarray}\label{g_lan_eq}
a_F P(z)+b_F P^3(z)-c_F \frac{d^2P(z)}{dz^2}=0, && 0 < z < L_1, \\
a_S \psi(z)+b_S \psi^3(z)-\frac{\hbar^2}{m e}
\frac{d^2\psi(z)}{dz^2}=0, && L_1 < z < L. \label{g_lan_eq2}
\end{eqnarray}
The extra surface terms appearing from the variation, give the set of the
boundary conditions at the right contact $z=L_1^+$
\begin{eqnarray}\label{bz_L1}
&& c_F\frac{dP(L_1^+)}{dz}-\frac{\gamma}{2} \psi^2(L_1^+)=0,\\
&& \frac{\hbar^2}{m e}\frac{d\psi(L_1^+)}{dz}+\gamma
\psi(L_1^+)P(L_1^+)=0,\nonumber
\end{eqnarray}
and at the left contact $z=0^-$ ($z=L^-$)
\begin{eqnarray}\label{bz_L}
&& c_F\frac{dP(0^-)}{dz}-\frac{\gamma}{2} \psi^2(L^-)=0,\\
&& \frac{\hbar^2}{m e}\frac{d\psi(L^-)}{dz}+\gamma \psi(L^-)P(0^-)=0.\nonumber
\end{eqnarray}
Setting $\gamma=0$ in (\ref{bz_L1}) and (\ref{bz_L}) results in ordinary free
boundary conditions (zero gradients of the order parameters) at the FE-S
contacts, which implies in fact the continuity of $P$ and $\psi$ near the
boundaries. However, a small finite coupling energy $\gamma \ne 0$ gives
non-zero gradients of $P$ and $\psi$
\begin{eqnarray}\label{cont}
&& \frac{dP_c}{dz} \sim \gamma\psi_c^2,\\
&& \frac{d\psi_c}{dz} \sim -\gamma \psi_c P_c, \nonumber
\end{eqnarray}
where $\psi_c$ and $P_c$ denote the corresponding contact values.

Consider the first expression (\ref{cont}) for $P$. As the right hand side is
always positive, we conclude that
\begin{equation}\label{cont_1}
  \frac{dP_c}{dz}>0.
\end{equation}
For $P>0$ this relation gives at the right contact $z=L_1$ an increase of $P$
due to the coupling with the surface charge, whereas at the left contact $z=0$,
$P$ decreases. Since for $P<0$ we have the similar behavior, we conclude that
independent of the direction of the polarization
\begin{equation}\label{p_inc}
P(z=L_1)>P(z=0).
\end{equation}

Let us analyze the second condition (\ref{cont}) for $\psi$ which in
distinction to the latter case depends on the direction of the polarization.
For $P>0$, the gradient $\frac{d\psi_c}{dz}<0$ and we obtain an increase of the
electronic pair density at $z=L_1$ and a lower density at $z=L$. Furthermore, a
switching of the polarization ($P<0$) results in $\frac{d\psi_c}{dz}>0$ and in
the opposite behavior of the charge density at the contacts.

Below we present the numerical results for the problem
(\ref{g_lan_eq})-(\ref{bz_L}) in the case when $T_c^F<T_c^S$, discussing in
limiting cases the analytical solutions for $P$ and $\psi$. Note that the case
$T_c^F<T_c^S$ is realized in complex insulating perovskite alloys such as BST
with high Sr content. The ferroelectric transition temperature in BST depends
on the Ba content $x$, approaching $T_c^F \approx 0$ for $x=0$ (SrTiO$_3$ is a
paraelectric material with no ferroelectric phase transition) and
$T_c^F=120^{\circ}$~C for $x=1$ (BaTiO$_3$). For instance, for $7.5\%$~Ba,
$T_c^F=60$~K \cite{lemanov} which lies already in the range of superconducting
transition temperatures of the HTS compounds.

\section{CASE STUDY: $T_c^F<T_c^S$ \label{phases}}

\begin{figure}[htbp]
\epsfxsize=8.5cm \centerline{\epsffile{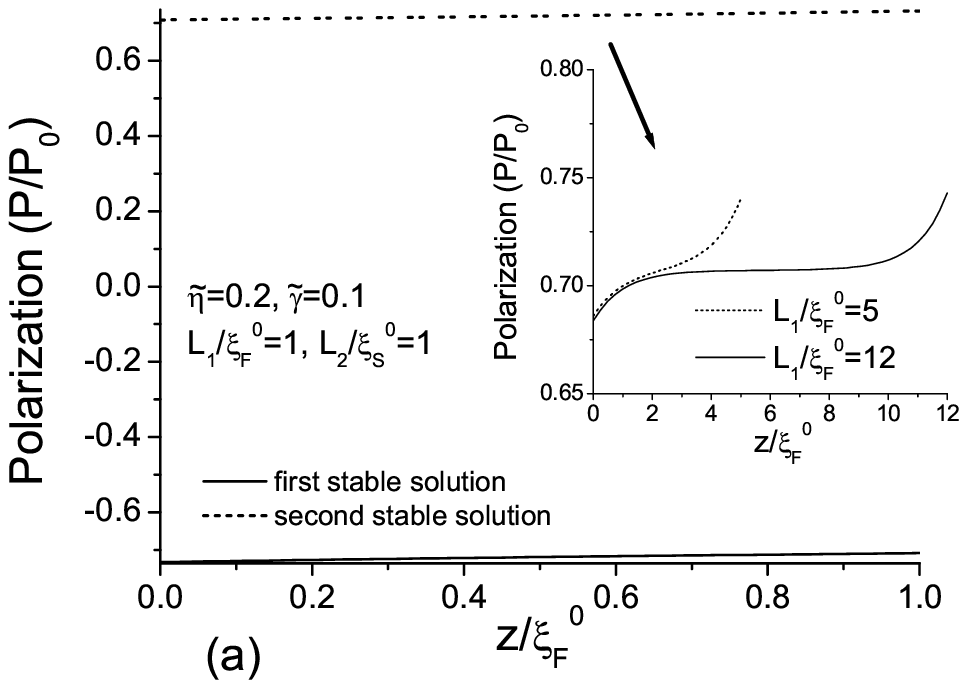}} \epsfxsize=8.5cm
\centerline{\epsffile{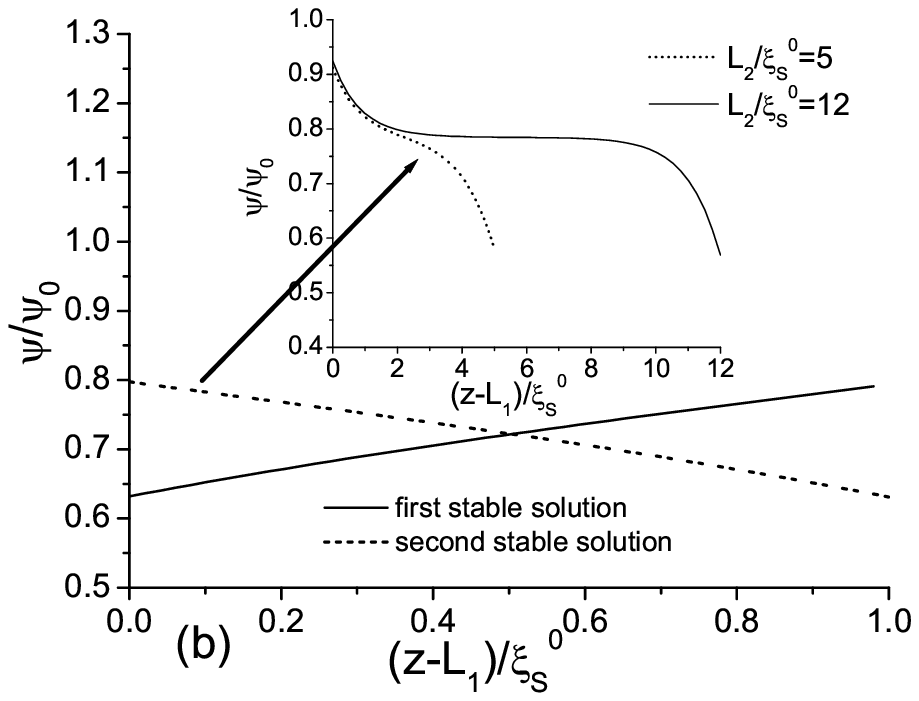}}
\caption{\label{fig1}Spatial profiles of (a) $P$ and (b) $\psi$ in the ordered
FE+SC phase at $\tau=0.5<\tau_c^F=1$ exhibiting the ferroelectric ordering in a
FE-film and the superconductivity in a S-film of moderate
thicknesses($L_1/\xi_F^0=1$ and $L_2/\xi_S^0=1$). The insets demonstrate the
corresponding profiles for thick layers.}
\end{figure}

We now present a brief overview of the results obtained for our
model (\ref{g_lan_eq})-(\ref{bz_L}) for different temperatures and
coupling $\gamma$. Fig.~\ref{fig1} shows spatial dependences of
$P$ and $\psi$ normalized by the zero-temperature values
$P_{0}=\sqrt{a_F^0/b_F}$ and $\psi_0=\sqrt{a_S^0/b_S}$ for the
following set of the parameters: $\tilde{\eta}=\eta \frac{\psi_0^2
\tilde{a}_S^0} {P_0^2}=0.2$ ($\eta=\xi_S^0/\xi_F^0$,
$\tilde{a}_S^0=a_S^0/a_F^0$) and $\tilde{\gamma}=\frac{\gamma}
{a_F^0 T_c^F}\frac{\psi_0^2}{\xi_F^0 P_0}=0.1$ at a low
temperature $\tau=T/T_c^F=0.5$, while $\tau_c^S=T_c^S/T_c^F=1.3$.
Here $\xi_S^0=\sqrt{\frac{\hbar^2}{2m e a_S^0 T_c^S}}$ is the
zero-temperature superconducting coherence length and the
correlation length for the FE-layer at $T=0$ is:
$\xi_F^0=\sqrt{\frac{c_F}{a_F^0 T_c^F}}$. Note that taking into
account the possible variation of the field penetration depth
(section~\ref{model}), we can estimate the range for the magnitude
of the dimensionless coupling $\tilde{\gamma}$: from $\sim
10^{-2}$ (for conventional superconductors) to $\sim 1$ (for HTS
films). For the films of moderate thickness ($L_1/\xi_F^0 = 1$ and
$L_2/\xi_S^0 =1$), one can observe a strong coupling effect
demonstrated in Fig.~\ref{fig1}. Similarly to the theory for bulk
ferroelectrics, the two possible directions of the polarization
can be stabilized in the FE-layer. As was already discussed in
section~\ref{Lan_ginz_eq} (equations (\ref{cont}) and
(\ref{cont_1})), $P$ increases at the right contact, $z=L_1$, and
lowers at the left one due to the coupling with the S-electrons
(Fig.~\ref{fig1}(a)). The important effect arising from this, is
the appearance of two stable solutions for $\psi$ in the adjacent
S-layer (Fig.~\ref{fig1}(b)). The first of them (with
$d\psi/dz<0$, $\psi(L_1)>\psi(L)$) corresponds to the positive
branch of $P$ whereas the second one ($\psi(L_1)<\psi(L)$) appears
due to the coupling with the negative $P$. Note that the regions
with the lower density $|\psi|^2$ (the depletion layer) and the
accumulation charge are located near the contacts
($L-\delta_S<z<L$) and ($L_1<z<L_1+\delta_S$) with $\delta_S \sim
\xi_S=\sqrt{|\frac{\hbar^2}{2m e a_S}|}$, whereas inside the thick
S-layer (Fig.~\ref{fig1}(b), inset), $\psi$ approaches its bulk
value $\psi_{b}=\sqrt{-\frac{a_S}{b_S}}$.

Similar interface effects with an increase of the FE-layer thickness are
demonstrated in the insets of Fig.~\ref{fig1}(a), where the regions of the
deviations of $P$ from its bulk value $P_{b}=\sqrt{-\frac{a_F}{b_F}}$ are
located within a distance $\delta_F \sim \xi_F=\sqrt{|\frac{c_F}{a_F}|}$ from
the interfaces.

The whole physical behavior becomes highly nontrivial for $T>T_c^F$ ($\tau>1$)
in the temperature range $\tau_c^F<\tau<\tau_c^S$.

\subsection{The phase with symmetric domains\label{enhanced_phase}}

\begin{figure}[htbp]
\epsfxsize=8.5cm \centerline{\epsffile{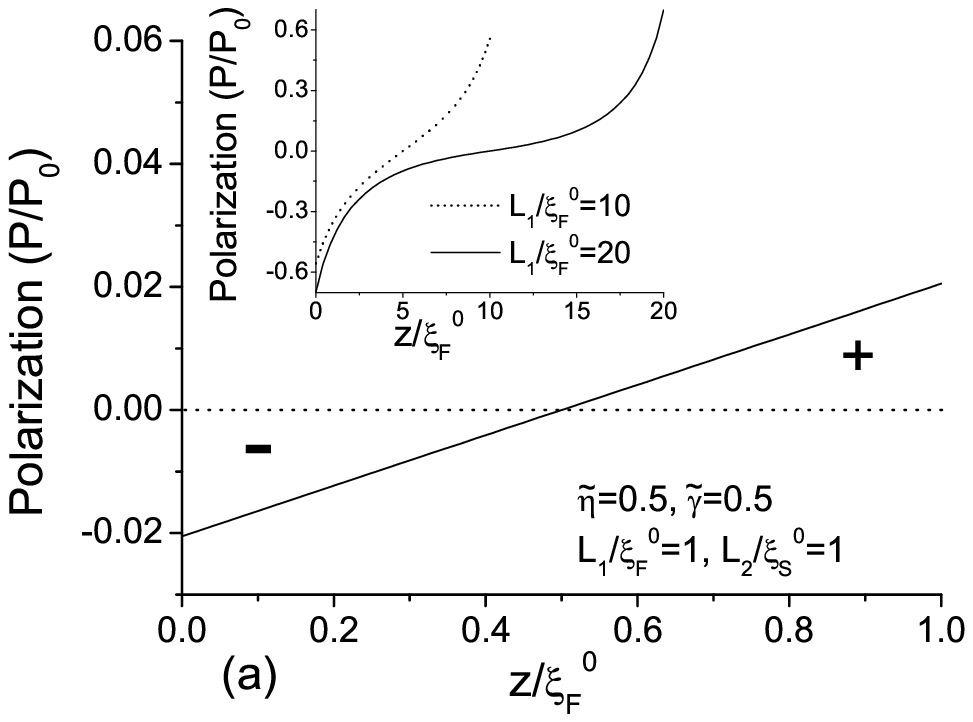}} \epsfxsize=8.5cm
\centerline{\epsffile{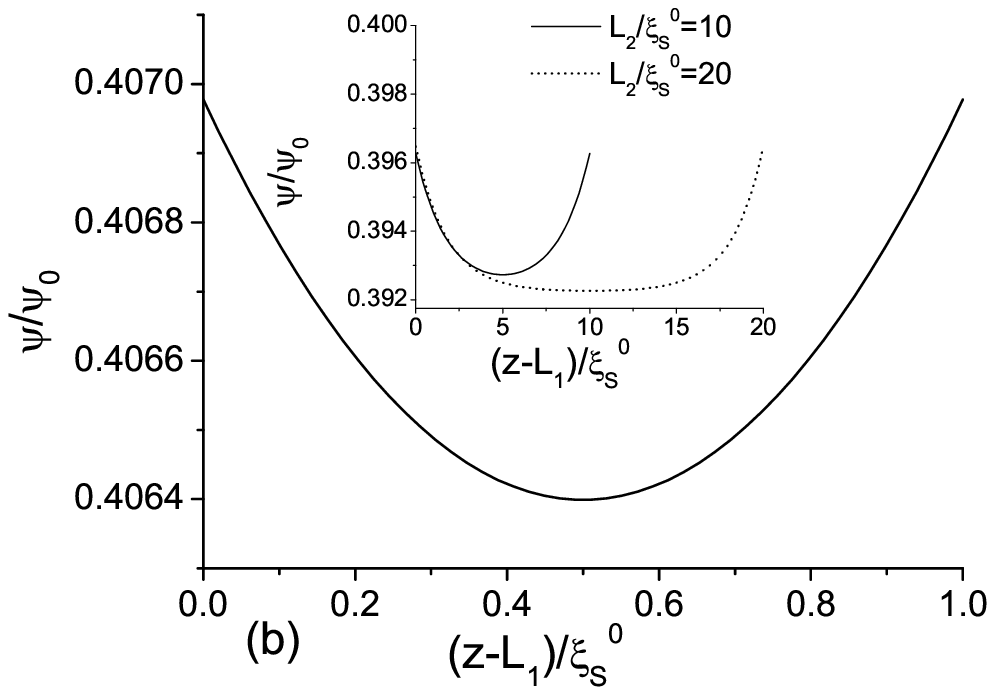}}
\caption{\label{fig2} Spatial profiles of (a) $P$ and (b) $\psi$ in the domain
phase at $\tau=1.1>\tau_c^F$ exhibiting the weak ferroelecity in a thin FE-film
and the superconductivity in a thin S-layer ($L_1/\xi_F^0=1$ and
$L_2/\xi_S^0=1$). The insets show the corresponding profiles for thick layers.}
\end{figure}

In contrast to the bulk, where $P=0$ for $T>T_c^F$, the coupling with the
interface charge leads to rather unusual behavior for $P$ and $\psi$ as shown
in Fig.~\ref{fig2} at $\tau=1.1$. The polarization in this case is enhanced,
and for $T_c^F<T<T_c^S$ and small $\tilde{\gamma}<1$ is given by

\begin{equation}\label{p_enh}
  P(z)=\frac{\gamma}{2c_F}\xi_F \frac{\sinh(z-\frac{L_1}{2})/\xi_F}{\cosh\frac{L_1}{2\xi_F}}
  |\psi(L^-)|^2,
\end{equation}
where the superconducting pair density at $z=L^-$
\begin{equation}\label{psi_L_enh}
  |\psi(L^-)|^2=\psi_b^2 \left[1+\frac{\gamma^2}{c_F b_S}\frac{\xi_F}{\xi_S}
  \tanh \frac{L_1}{2\xi_F}\cot \frac{L_2}{2\xi_S}\right]+\mathcal{O}(\gamma^4).
\end{equation}

As follows from (\ref{p_enh}), in this state a non-zero $P$ exists only due to
the superconducting electrons, because setting $\gamma=0$ or $\psi(L)=0$
immediately yields $P=0$.

For the thin FE-films with $\frac{L_1}{\xi_F}< 1$ or close to the bulk
transition temperature (where $\xi_F \rightarrow \infty$), the effect of the
contacts is particularly pronounced and the polarization is linear in $z$
\begin{equation}\label{thin_P}
  P(z)=\frac{\gamma}{2c_F} (z-\frac{L_1}{2})|\psi(L)|^2.
\end{equation}
However, as the FE-layer becomes thicker ($\frac{L_1}{\xi_F}\gg 1$), the
polarization deviates from zero only near the interfaces, whereas inside the
layer $P \rightarrow 0$. For example, for $z \approx 0$
\begin{equation}\label{p_enh_0}
  P \sim -\gamma \xi_F \psi_b^2 e^{-z/\xi_F}
\end{equation}
and decays exponentially as one moves towards the interior of the layer. Near
the right contact $z \approx L_1$, the behavior is similar
\begin{equation}\label{p_enh_L1}
  P \sim \gamma \xi_F \psi_b^2 e^{-(L_1-z)/\xi_F}.
\end{equation}
Note that the polarization is antisymmetric with respect to the center of the
FE-layer $z={L_1}/{2}$: $P(z)=-P(L_1-z)$, forming two $P$-domains: with a
negative $P$ for $0 \le z \le L_1/2$, and a positive $P$ for $L_1/2 \le z \le
L_1$, as shown in Fig.~\ref{fig2}(a). The center of the layer, where
$P(z=L_1/2)=0$, corresponds to the domain wall.

Consider now the behavior of the S-layer. For thin S-films ($\frac{L_2}{\xi_S}
< 1$, $L_2 \ne 0$) in the vicinity of the $T_c^F$ ($T_c^F<T<T_c^S$), and for a
weak coupling $\tilde{\gamma}<1$, the superconducting order parameter is given
by
\begin{equation}\label{psi_thin}
  \psi(z)=\psi_b \left[1+\frac{\gamma^2}{2c_F b_S \eta} \tanh \frac{L_1}{2\xi_F}
   \frac{\cos \frac{z-L_1-L_2/2}{\xi_S}}{\sin\frac{L_2}{2\xi_S}} \right]+\mathcal{O}(\gamma^4).
\end{equation}
In contrast to the antisymmetric behavior of $P$ in the domain-type phase,
$\psi(z)$ is symmetric about the center of the S-layer $z=L_1+L_2/2$:
$\psi(L_1+z)=\psi(L-z)$ as can be seen in Fig.~\ref{fig2}(b). We note that for
the case of thin S- or FE-films ($L_1/\xi_F \ll 1$ or $L_2/\xi_S \ll 1$) the
effect of fluctuations is expected to be essentially important. This can be
studied by including the additional higher-order terms into the free-energy
functional (\ref{f_ferr})-(\ref{f_s}) that will be a subject of separate work.
From (\ref{psi_thin}) it follows immediately that as $\gamma \rightarrow 0$ or
$\eta=\frac{\xi_S}{\xi_F} \rightarrow \infty$, $\psi(z)$ approaches its bulk
value $\psi_b$. Moreover, we obtain the bulk behavior in (\ref{psi_thin}) also
when the adjacent FE-film becomes thin, so that $L_1 \ll \xi_F$. Similarly to
the FE-layer, in a thick S-layer the influence of the interfaces is remarkable
only near the contacts (Fig.~\ref{fig2}(b), inset).

To gain a better understanding of the nature of the domain phase, we determine
its free energy which can be obtained by the substitution of $P$ and $\psi$
from (\ref{p_enh}) and (\ref{psi_thin}) into (\ref{f_ferr}), (\ref{f_s}) and
(\ref{f_fe_s})
\begin{equation}\label{f_enh}
  F_{dmn}=-\frac{\gamma^2}{4c_F} \xi_F \psi_b^4 \tanh \frac{L_1}{2\xi_F} -L_2 \frac{b_S}{4}\psi_b^4
  +\mathcal{O}(\gamma^4).
\end{equation}
Note that since for the free (non-coupled) FE-layer $F_{FE}^0=0$ for $T>T_c^F$,
the total free energy for the non-coupled FE+S multilayers for $T_c^F<T<T_c^S$
is given by
\begin{equation}\label{f_nc0}
  F_{FE+S}^0=-L_2 \frac{b_S}{4}\psi_b^4.
\end{equation}
Therefore, the domain state always has an energy gain due to a negative
difference between the energies in these two phases
\begin{equation}\label{df_enh_0}
  \Delta F_{dmn}=F_{dmn}-F_{FE+S}^0=-\frac{\gamma^2}{4c_F} \xi_F \psi_b^4 \tanh \frac{L_1}{2\xi_F}
  +\mathcal{O}(\gamma^4).
\end{equation}
The latter result supports our conclusion that a non-zero coupling with $\psi
\ne 0$ is the main reason for the stabilization of the domain state. We note
also that as $\gamma \rightarrow 0$, we obtain that the energy gain $\Delta
F_{dmn} \rightarrow 0$, the polarization $P \rightarrow 0$, and hence the
domain phase transforms in fact into an ordinary paraelectric state:
\begin{eqnarray*}
  T_c^{F,\gamma} \rightarrow T_c^F, \quad {\rm as}\quad \gamma \rightarrow 0.
\end{eqnarray*}
Thus in the following analysis we define $T_c^{F,\gamma}$ as the temperature of
the transition between the ordered FE+SC phase with the bulk-type $P \ne 0$,
and the domain-type phase, which occurs on heating.

\subsection{Strong coupling behavior}

\begin{figure}[htbp]
\epsfxsize=8.5cm \centerline{\epsffile{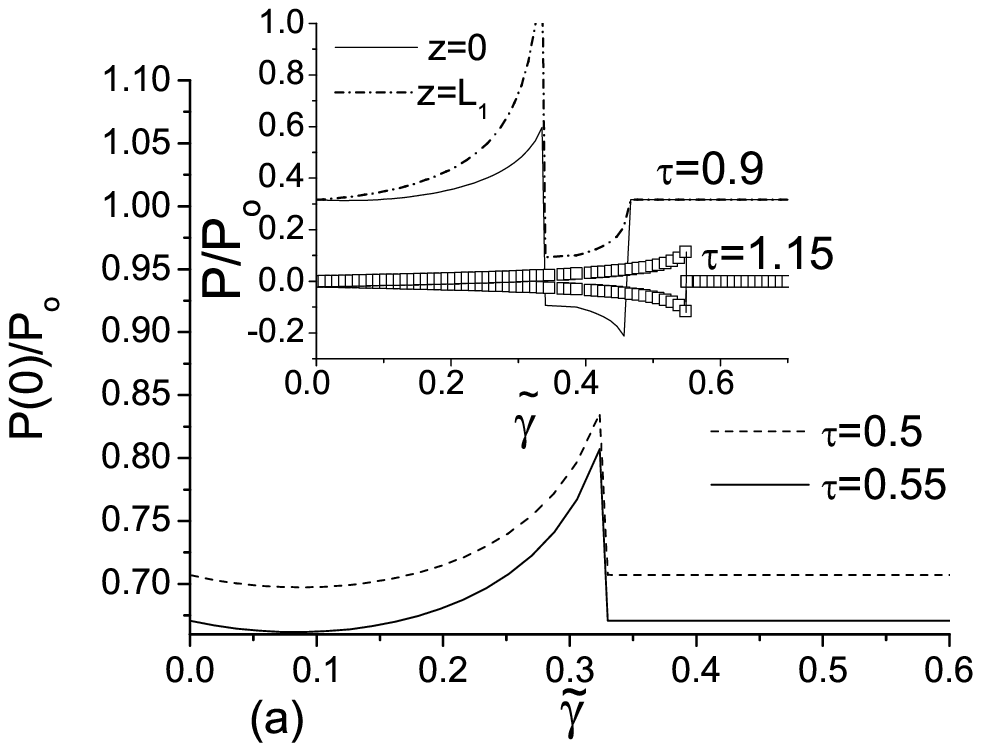}} \epsfxsize=8.5cm
\centerline{\epsffile{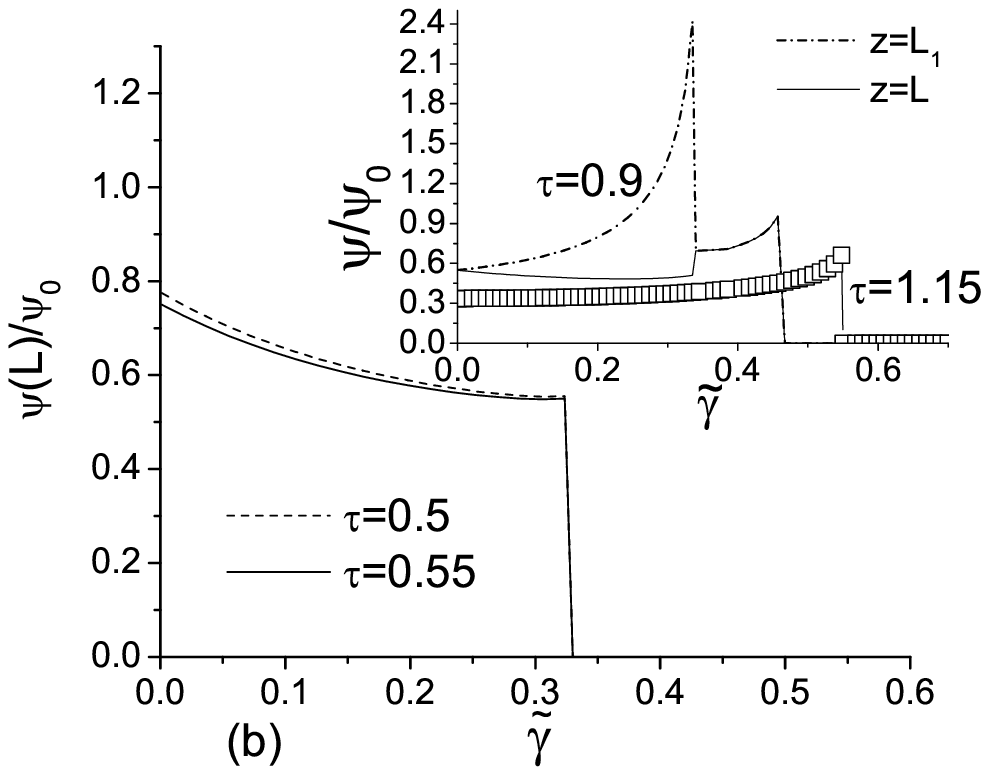}}
\caption{\label{fig3}(a) Polarization in the FE-film of thickness
$L_1/\xi_F^0=2$ and (b) $\psi$ at $z=L$ in the S-film of thickness
$L_2/\xi_S^0=2$ vs the magnitude of the coupling $\tilde{\gamma}$, exhibiting a
transition from the FE+SC phase to the FE-state with suppressed
superconductivity. Here $\tilde{\eta}=0.5$. The insets show the $P$ and $\psi$
behavior at the FE-S contacts with increasing $\tilde{\gamma}$ for $\tau=0.9$
(solid curves and dashes) and $\tau=1.15$ (open squares) with the stabilization
of the domain state for $\tilde{\gamma}<\tilde{\gamma}_c^{e}$.}
\end{figure}

As results from the previous section, generally a weak coupling $\tilde{\gamma}
\ll 1$ at the contacts produces symmetric domain phase with a non-zero
polarization for $T_c^F<T<T_c^S$. We demonstrate how this behavior changes as
the coupling becomes stronger, i.e., with the increasing $\tilde{\gamma}$. In
Fig.~\ref{fig3}, we plot the order parameters vs $\tilde{\gamma}$ for layers of
thicknesses $L_1/\xi_F^0=2$ and $L_2/\xi_S^0=2$. At lower $\tau$ ($\tau=0.5$
and $\tau=0.55$), the behavior of $\psi$ and $P$ rapidly changes at
$\tilde{\gamma}\approx 0.33$, corresponding to the transition to the normal
(non-superconducting) state ($\psi=0$). In this state, $P=P_{b}=const$ does not
depend on $z$ because setting $\psi=0$ in (\ref{bz_L}) and (\ref{bz_L1}) gives
the free boundary conditions. The suppression of the superconductivity due to
the strong contact with polarization resembles the effect of an external field,
impurities and other pair-breaking factors on superconductors\cite{tinkham}.
Like these factors, the polarization at the boundaries with the S-layer
supplies additional energy (\ref{f_fe_s}) for the breaking of electron pairs in
the strong-coupling regime.

From the point of view of the Landau-Ginzburg functional (\ref{f_ferr}),
(\ref{f_s}), (\ref{f_fe_s}), an increase of $\gamma$ induces larger values
of the fourth-order fluctuation terms as well as the third terms in (\ref{f_ferr}) and
(\ref{f_s}) describing the spatial inhomogeneities, and hence drives a
transition to the homogeneous state with $\psi=0$ which has the lower energy.
To explain qualitatively the effect of $\gamma$, we consider for simplicity
only the left FE-S contact with the energy $\frac{\gamma}{2}P(0^-)\psi^2(L^-)$
assuming at the right contact the free boundary conditions. At low $T$ the
expression for the free energy  in the ordered FE+S-phase is given by
\begin{equation}\label{f_0}
  F_{FE+S}\approx -\frac{1}{4}[P_b^4b_F L_1+\psi_b^4b_S L_2]+\frac{\gamma}{2}P_b \psi_b^2.
\end{equation}
The critical value for $\gamma$ can be found by equating (\ref{f_0}) and the
free energy for the state with the suppressed superconductivity
\begin{equation}\label{f_suppr}
  F_{suppr}=-\frac{1}{4}P_b^4b_F L_1,
\end{equation}
which yields
\begin{equation}\label{gamma_c}
  \tilde{\gamma}_c=\frac{\psi_b^2}{2P_b^2} \frac{\tilde{a}_0^S}{T_c^F} \eta
  \frac{L_2}{\xi_S^0}.
\end{equation}
From (\ref{gamma_c}) it follows, that as the S-layer becomes thicker, a stronger critical
coupling is required to destroy the superconducting state. In contrast, a
larger polarization $P$ in Eq.~(\ref{gamma_c}) reduces $\tilde{\gamma}_c$, which
supports our interpretation of the analogy with other external field pair
breaking effects.

Consider now the $\gamma$-dependence at higher temperatures shown in the insets
of Fig.~\ref{fig3}(a) and Fig.~\ref{fig3}(b) slightly below ($\tau=0.9$) and
above ($\tau=1.15$, where the system  for small $\tilde{\gamma}$ is in the domain
phase) the bulk value $\tau_c^F=1$. Like at low temperatures,
the superconductivity of the domain phase ($\tau=1.15$) is destroyed
when approaching the critical value $\tilde{\gamma}_c^{e} \approx 0.55$. We note that
since $\psi=0$ for $\tilde{\gamma}>\tilde{\gamma}_c^{e}$, the polarization, which is
enhanced here only due to
the coupling with $\psi$, vanishes and we obtain the combined
paraelectric+suppressed superconducting ($P=0$, $\psi=0$) phase for
$\tau>\tau_c^F$. We emphasize that in contrast to the bulk ferroelectrics
exhibiting $P=0$ in this temperature range, in our coupled FE-layer the zero
polarization is only possible due to the strong coupling effect depressing the
source for the enhancement of $P$.

The behavior for large $\tilde{\gamma}$ can be even more complicated as demonstrated in
the inset of Fig.~\ref{fig3} at the temperature $\tau=0.9$ which is below but
close to $\tau_c^F$ (Fig.~\ref{fig3}(a)). Despite $\tau<\tau_c^F$, the
$\gamma$-dependence of $P$ and $\psi$ clearly shows the stabilization of the
domain state with $P(0)=-P(L_1)$ and $\psi(L_1)=\psi(L)$ in the range
($0.34<\tilde{\gamma}<0.46$), and only with the further $\tilde{\gamma}$ increase
($\tilde{\gamma}>0.46$), the transition to the ($P=P_{b}$, $\psi=0$)-state. To
analyze more comprehensively the behavior for different $\tilde{\gamma}$, we
discuss in the next section the phase diagrams ($\tilde{\gamma}$, $\tau$) plotted
for the different FE- and S-layer thicknesses.

\section{PHASE DIAGRAMS ($\tilde{\gamma}$, $\tau$)\label{diagrams}}

\begin{figure}[htbp]
\epsfxsize=8.cm \centerline{\epsffile{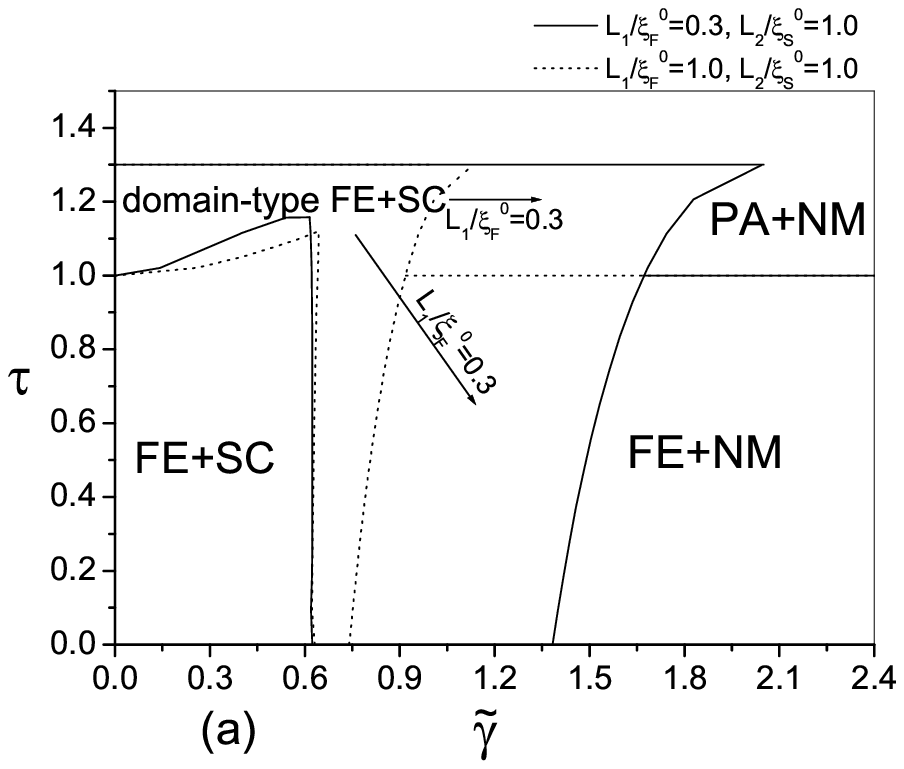}} \epsfxsize=8.cm
\centerline{\epsffile{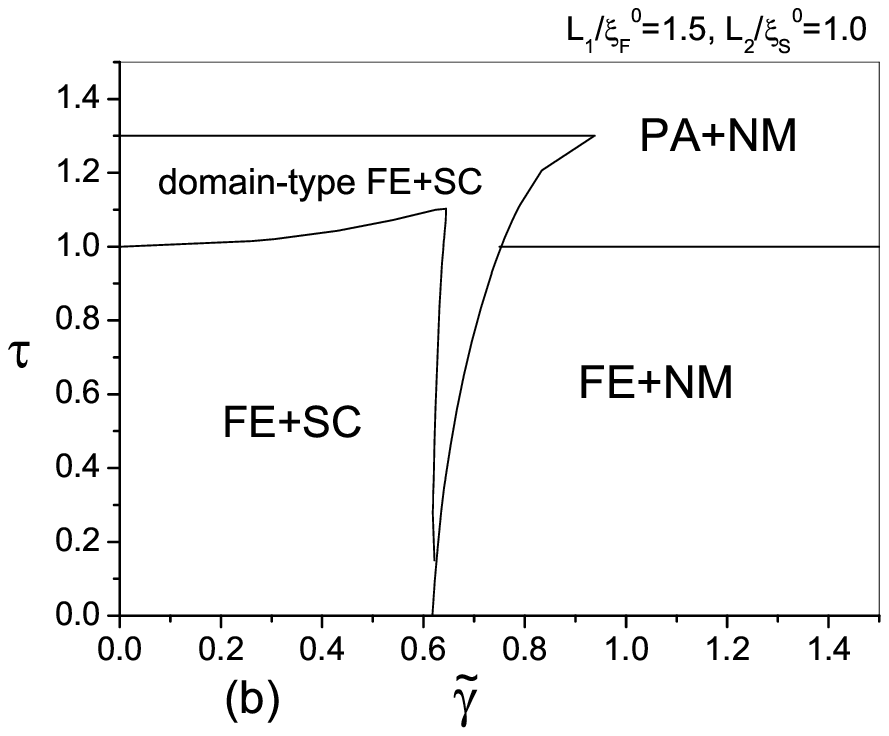}} \epsfxsize=8.cm
\centerline{\epsffile{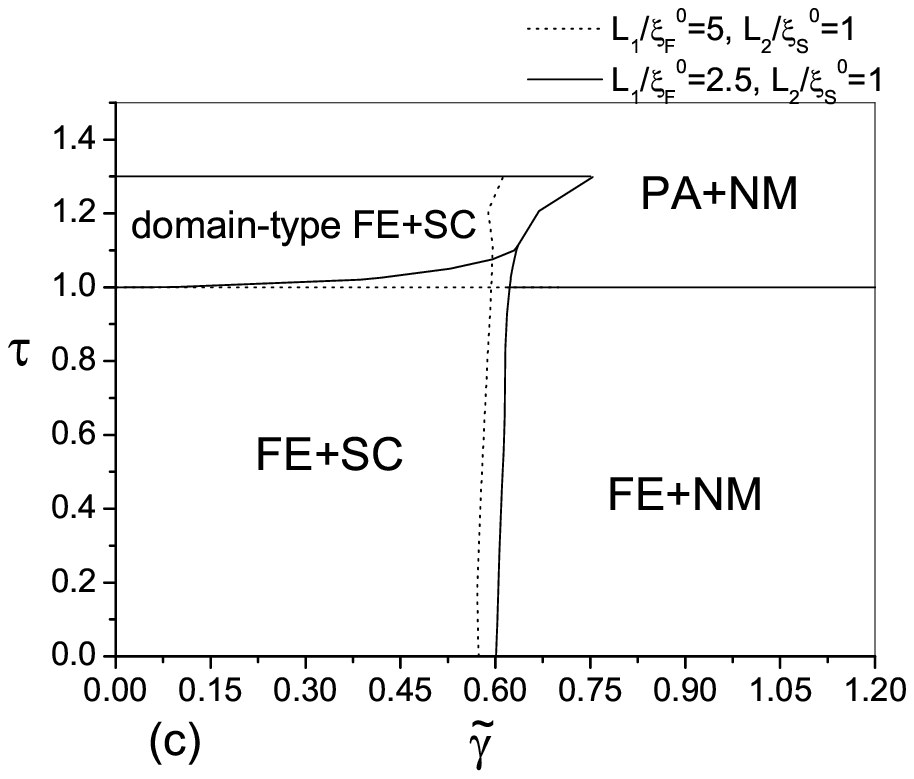}}
\caption{\label{fig4}Phase diagrams ($\tilde{\gamma}$, $\tau$) for a compound
containing S-films of thickness $L_2/\xi_S^0=1.0$ coupled to the FE-layers of
thickness: (a) $L_1/\xi_F^0=0.3$ and $L_1/\xi_F^0=1.0$ (thin FE-films); (b)
$L_1/\xi_F^0=1.5$ (moderately thick FE-film) and (c) $L_1/\xi_F^0=2.5, 5.0$
(thick FE-layer). Here $\tilde{\eta}=0.5$.}
\end{figure}

\begin{figure}[htbp]
\epsfxsize=8.cm \centerline{\epsffile{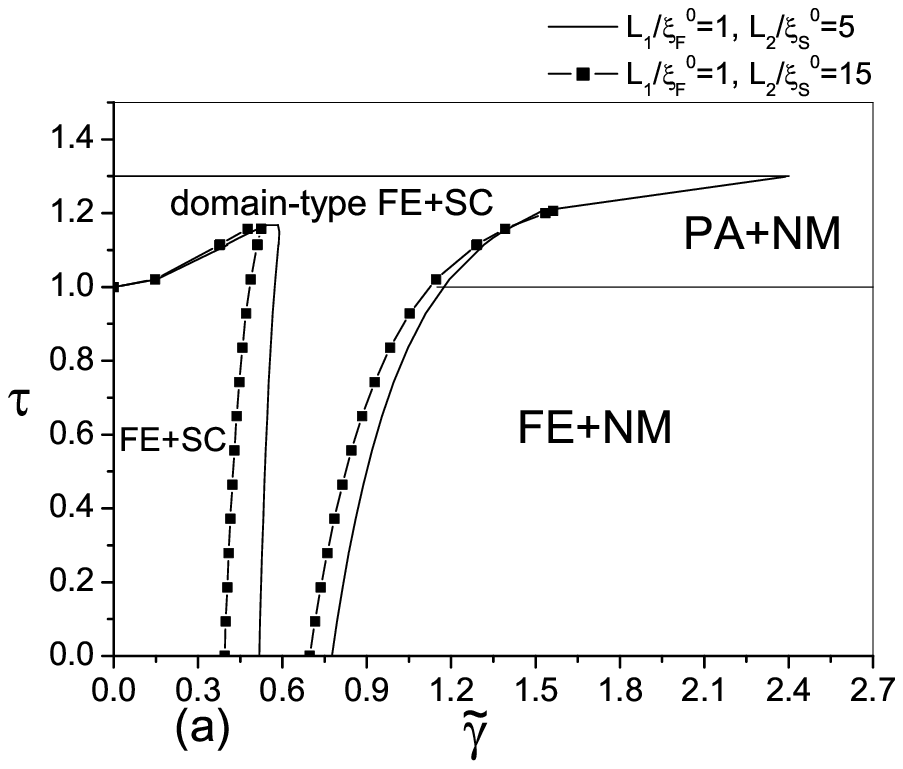}} \epsfxsize=8.cm
\centerline{\epsffile{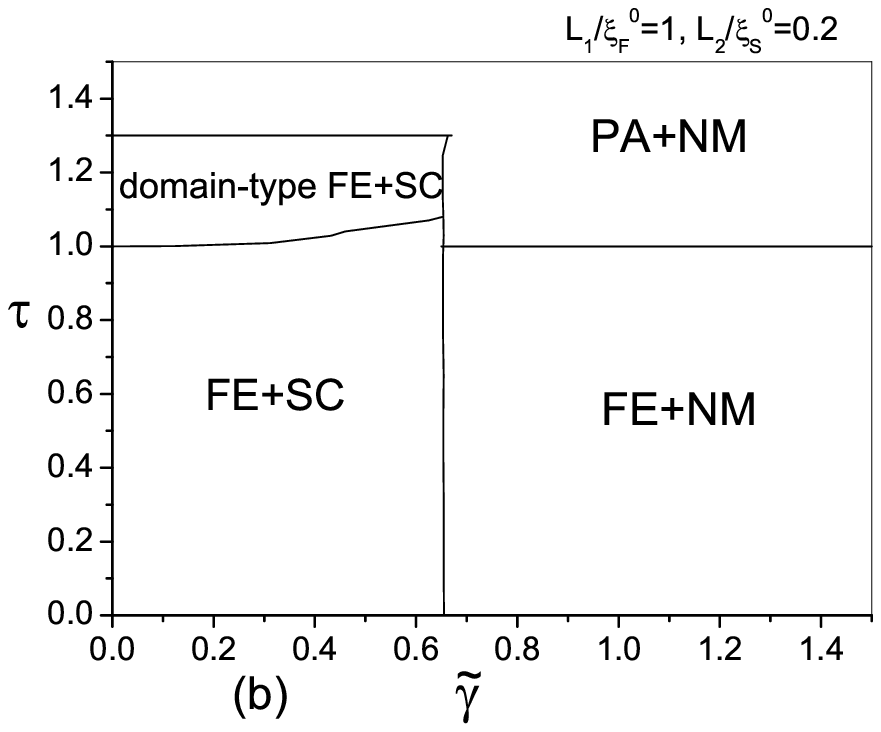}}
\caption{\label{fig5}Phase diagrams ($\tilde{\gamma}$, $\tau$) for a compound
containing FE-films of thickness $L_1/\xi_F^0=1.0$ coupled to S-layers of
thickness: (a) $L_2/\xi_S^0=5.0$ and $L_2/\xi_S^0=15.0$ (thick S-layer) and (b)
$L_2/\xi_S^0=0.2$ (thin S-film). Here $\tilde{\eta}=0.5$.}
\end{figure}

The phase diagrams ($\tilde{\gamma}$, $\tau$) plotted in Fig.~\ref{fig4} and
Fig.~\ref{fig5} show the regions of stability for different phases discussed in
section~\ref{phases}. As has been revealed in the foregoing analysis, the
following phases can exist in our periodical multi-layer system: (i) the ordered
bulk-type FE+SC phase with the nonzero spontaneous polarization in the FE-layer and
the superconductivity in the S-layer; (ii) the superconductivity in the S-layer coexisting
with the ferroelectric domain state exhibiting the symmetric (negative and positive)
polarization domains due to the FE-S contacts (so called domain-type phase);
(iii) the phase with the ferroelectricity and the suppressed superconductivity
(FE+NM); (iv) the high-temperature paraelectric (in the FE-layer)+normal state
(in the S-layer).

In Fig.~\ref{fig4} we plot a set of the diagrams for the S-film of moderate
thickness $L_2/\xi_S^0=1$ coupled to the FE-layers of different thickness:
(a) a thin FE-film, $L_1/\xi_F^0=0.3, 1.0$, (b) a FE-layer of moderate thickness
$L_1/\xi_F^0=1.5$, and (c) a thick FE-layer with $L_1/\xi_F^0=2.5, 5.0$. We
associate in the following analysis the variation of the FE-thickness as
related to that of the S-layer, to the ratio
\begin{eqnarray}\label{zeta}
\zeta=\frac{L_1}{L_2}=\eta^{-1} \frac{L_1/\xi_F^0}{L_2/\xi_S^0},
\end{eqnarray}
where the ratio between the superconductor and the ferroelectric
coherence lengths $\eta$ ranges from $\sim 0.05$ (for HTS
compounds) to $\sim 10$ (conventional superconductors), taking the
typical value for $\xi_F^0\approx 10$~nm
(Ref.~\onlinecite{lines}). As $\zeta$ decreases from $5/\eta$
(Fig.~\ref{fig4}(c)) to $0.3/\eta$ (Fig.~\ref{fig4}(a)), the area
of the domain-type phase drastically expands towards larger
$\tilde{\gamma}$ and lower $T$ and finally approaches even the
low-temperature range as represented by arrows in
Fig.~\ref{fig4}(a). The expansion of the domain phase is
accompanied by the simultaneous decrease of the region of the
FE+NM(suppressed)-phase. We note that in the system with a thicker
FE-layer ($L_1/\xi_F^0=1$, Fig.~\ref{fig4}(a)), the transition to
the suppressed state occurs at a significantly lower
$\gamma_c^{e}$ ($\tilde{\gamma}_c^{e} \approx 0.9$) as compared to
a thin FE-film (case $L_1/\xi_F^0=0.3$ in Fig.~\ref{fig4}(a)) with
$\tilde{\gamma}_c^{e} \approx 1.8$ ($\tau=1.1$), which clearly
demonstrates the crucial role of the ferroelectric layer in the
suppression of the superconductivity. These modifications of the
phase diagram can be explained by the competition between {\it the
first tendency} to destroy the superconductivity prevailing for
the coupling with thicker FE-layers, and {\it the second tendency}
to retain the superconducting state and stabilize $P$-domains
dominating in systems with thicker S-layers. As is demonstrated in
Fig.~\ref{fig4}, the decrease of $\zeta$ results in a dominant
role of the S-layer and, as a consequence, in the enhancement of
the region with $\psi \ne 0$ at lower $T<T_c^F$ (which in turn
induces $P$-domains in the FE-layer). Since in the HTS-films
$\eta$ is about three orders of magnitude smaller than in
conventional superconductors, one might expect from (\ref{zeta})
that even a slight variation of the S- or FE-film thickness leads
to a dramatic modification of the phase diagram and, consequently,
of the stable state of this type of compounds.

Fig.~\ref{fig5} shows the behavior of the system with a fixed FE-thickness
$L_1/\xi_F^0=1$ and with a S-thickness $L_2/\xi_S^0$ varying from $15$
(Fig.~\ref{fig5}(a)) to $0.2$ (Fig.~\ref{fig5}(b)). As in the previous case, we
find here the same tendency of expansion of the domain phase with decreasing $\zeta$.
We also bring to attention that the destruction of the superconducting state
demonstrated here has been observed in the measurements of the resistivity in the film
YBCO/BaTiO$_3$ composites \cite{gajevskis}.

\subsection{Comparison with the microscopic model of Ref.~\onlinecite{pavlenko}}

As has already been mentioned in section~\ref{model}, when the ferroelectric critical
temperature $T_c^F$ is close to $T_c^S$, the
phenomenological approach proposed in this work, has common key aspects with the
microscopic description proposed in Ref.~\onlinecite{pavlenko}. However, while
the latter model describes the superconducting layer as two boundary 2-D planes
exhibiting BCS-superconductivity, the present approach allows us to study the behavior
inside the thick S-layers. Despite the fact that the microscopic BCS-model of
Ref.~\onlinecite{pavlenko} does not consider directly the internal processes in the
superconducting layer and treats the inner part of the layer as a source for the
electron transfer
towards (away from) the boundary S-planes, the following key results appear in both approaches:\\
(i) the enhancement of the ferroelectric
polarization with a formation of positive and negative $P$-domains above the ferroelectric
transition temperature as well as for large $\gamma$ and low temperatures;\\
(ii) the suppression of the superconductivity for the strong FE-S coupling
due to the ferroelectric field-type effect.\\
Thus, the basic structure of the phase diagrams ($\gamma$, $T$) together with
the behavior of the system in different observed phases discussed in
Ref.~\onlinecite{pavlenko}, also retains its fundamental features in the
present phenomenological approach.

\subsection{Dependence of $T_c^{F,\gamma}$ on the FE- and S-layer thicknesses}

The behavior of the different phases discussed above with the variation of the FE-
and S-thicknesses results in different types (from monotonous to complex
nonmonotonic) of the corresponding $T_c^{F,\gamma}$ dependences.

\begin{figure}[htbp]
\epsfxsize=9.cm \centerline{\epsffile{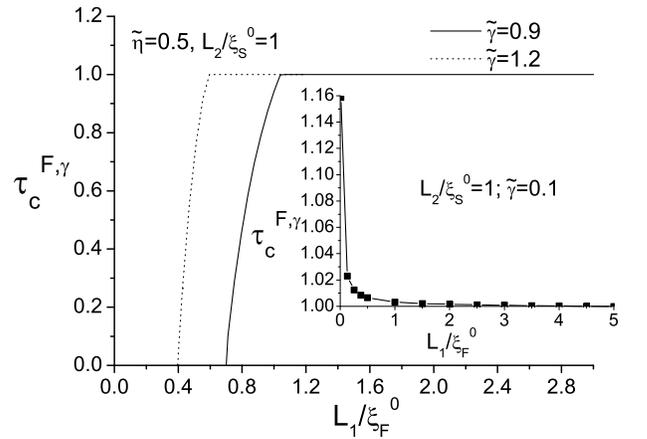}} \epsfxsize=9.cm
\caption{\label{fig6}Ferroelectric transition temperature vs the FE-layer
thickness (for a S-film with $L_2/\xi_S^0=1.0$) for $\tilde{\eta}=0.5$ in the
strong coupling regime. The inset shows $\tau_c^{F,\gamma}$ in the
weak-coupling regime for $\tilde{\gamma}=0.1$.}
\end{figure}

\begin{figure}[htbp]
\epsfxsize=9.cm \centerline{\epsffile{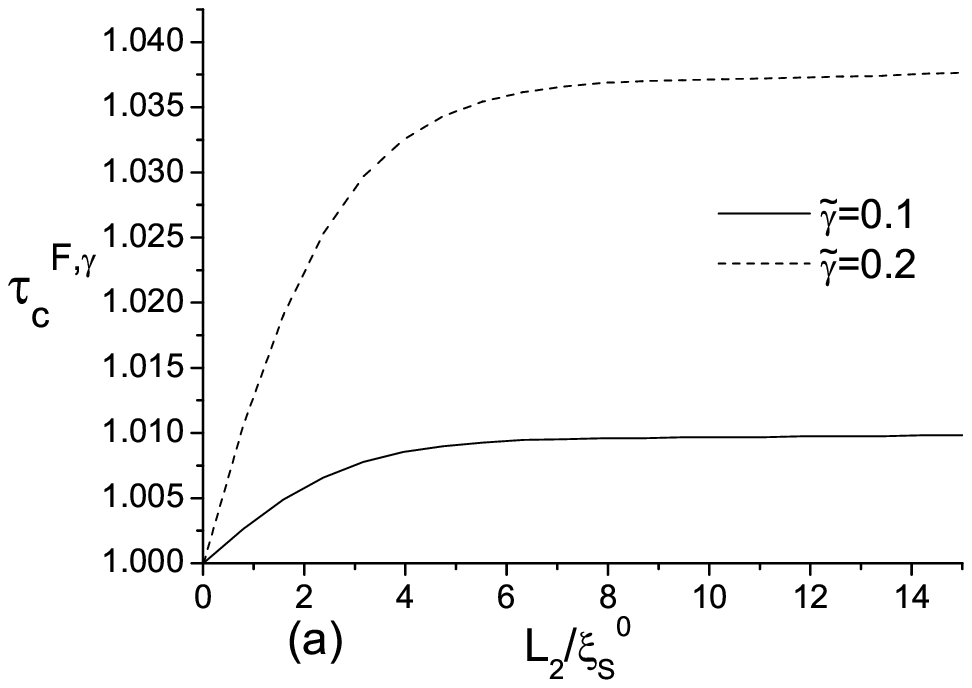}} \epsfxsize=9.cm
\centerline{\epsffile{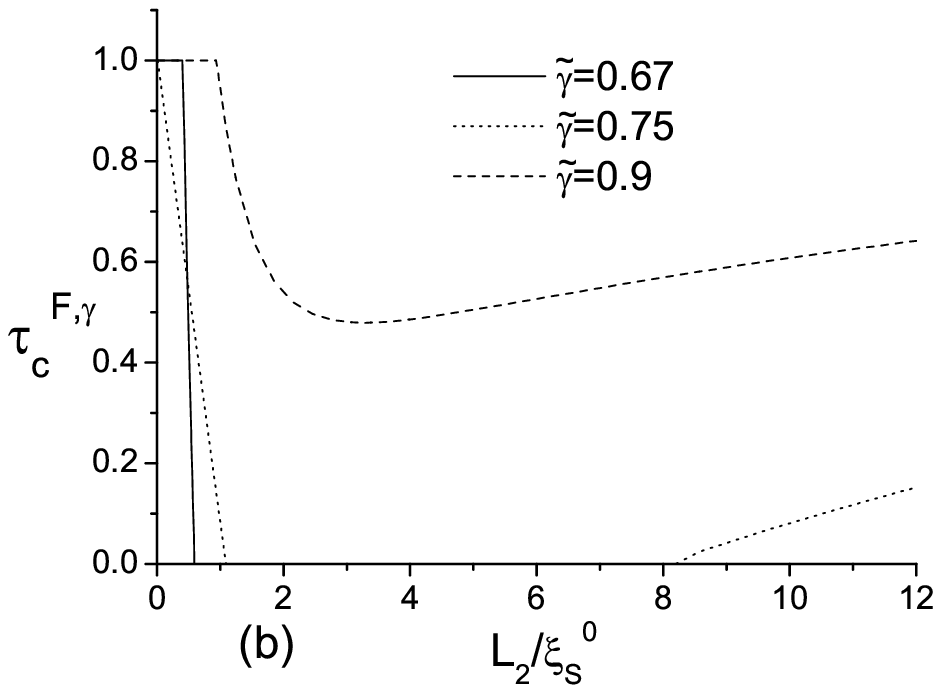}}
\caption{\label{fig7}Ferroelectric transition temperature vs the S-layer
thickness (for a FE-film with $L_1/\xi_F^0=1.0$) for $\tilde{\eta}=0.5$ and
different values of coupling strength $\tilde{\gamma}$: (a) weak coupling
regime and (b) strong coupling regime.}
\end{figure}

Since the system properties drastically change for the large coupling
$\tilde{\gamma}>\tilde{\gamma}_c$, the $L_1$- and $L_2$-dependence of $T_c^{F,\gamma}$
is also qualitatively different in the weak
($\tilde{\gamma}<\tilde{\gamma}_c$) and in the strong coupling regime
($\tilde{\gamma}>\tilde{\gamma}_c$). To illustrate this, in Fig.~\ref{fig6} and
Fig.~\ref{fig7} we plot several curves for various values of $\tilde{\gamma}$.
Specifically, Fig.~\ref{fig6}(inset) and Fig.~\ref{fig7}(a) show the case
$\tilde{\gamma}<\tilde{\gamma}_c$, whereas
$\tilde{\gamma}>\tilde{\gamma}_c$ is represented by Fig.~\ref{fig6}($\tilde{\gamma}\ge
0.9$) and Fig.~\ref{fig7}(b).

Depending on the value of $\tilde{\gamma}$, we can obtain the following characteristic
types of $T_c^{F,\gamma}(L_1)$ and $T_c^{F,\gamma}(L_2)$ behavior:\\
(1) {\it for $\tilde{\gamma}<\tilde{\gamma}_c$}, $\tau_c^{F,\gamma}=T_c^{F,\gamma}/T_c^F$
decays monotonically with increasing $L_1$ to the bulk value $\tau_c^F=1$
(Fig.~\ref{fig6}, $\tilde{\gamma}=0.1$ in the inset). We note that in contrast to the
model in Ref.~\onlinecite{pavlenko} where $T_c^{F,\gamma}$ lies slightly below
$T_c^{F,bulk}$, the coupling with a S-layer of a finite thickness produces here the
enhancement of the bulk-type FE+SC-phase. This corresponds to the increase of $T_c^{F,\gamma}$ ($T_c^F<T_c^{F,\gamma}<T_c^S$) for $\gamma \ne 0$ ($\tilde{\gamma} <\tilde{\gamma}_c$), observed also in the phase diagrams in Figs.~\ref{fig4} and \ref{fig5}.
As a result, we also obtain that:\\
(2) {\it for $\tilde{\gamma}<\tilde{\gamma}_c$}, the increase of $L_2$ produces a
monotonic increase of $\tau_c^{F,\gamma}$, with finally approaching a saturation value
which is the higher the larger $\tilde{\gamma}$ (Fig.~\ref{fig7}(a)).\\
(3) {\it For $\tilde{\gamma}>\tilde{\gamma}_c$}, the complex behavior of
$\tau_c^{F,\gamma}$ vs $L_1$ shown in Fig.~\ref{fig6}($\tilde{\gamma} \ge
0.9$), is governed by a decrease of the domain phase-region as $L_1$ increases
(see the plots in Fig.~\ref{fig4}). First, for thin FE-films, the system is in
the domain phase already for $T \sim 0$. As $L_1/\xi_F^0$ increases, the area
of the domain state shrinks towards higher $T$  and finally disappears, which
results in an increase of $\tau_c^{F,\gamma}$ up to $\tau_c^F=1$ related to the
transition from the FE+NM(suppressed) to the paraelectric state (see the
corresponding diagram in Fig.~\ref{fig4}(c)). Note that for a small thickness
of the FE-film ($L_1/\xi_F^0 \sim 0.5-1$ in Fig.~\ref{fig6}), the behavior of
$\tau_c^{F,\gamma}$ reflects a "crossover" between these two different
ferroelectric transition types, namely, between the transition from the
FE+NM(suppressed) to the domain state for the thin FE-films, and the transition
from the FE+NM to the paraelectric phase for the thick FE-layers. This effect
is also seen in the large $\tilde{\gamma}$-region of the diagrams plotted
in Fig.~\ref{fig4}.\\
(4) Fig.~\ref{fig7}(b) shows a nonmonotonic dependence of $\tau_c^{F,\gamma}$
on $L_2$ {\it for $\tilde{\gamma}>\tilde{\gamma}_c$}. As in the $L_1$-dependences, for a
given value of $\tilde{\gamma}$ at a moderate thickness $L_1/\xi_S^0 \approx 1$, we
find similar, but reversed "crossover" between the two above-mentioned
different transition types. Specifically, while for thin S-films, the increase
of $T$ results in the transition from the FE+NM(suppressed) to the paraelectric
state (at $\tau_c^{F,\gamma}=1$), the increase of the S-thickness induces the
enhancement of the domain-type phase, leading
to a lower $\tau_c^{F,\gamma}$ (which now corresponds to the transition between
the FE+NM and the domain phases). See the evolution of the phase diagram
topology as $L_2/\xi_S^0$ varies, shown in Fig.~\ref{fig5}. Furthermore, we
note that for the larger $L_2/\xi_S^0>0.7$ the transition from the
FE+NM- to the symmetric domain-state exhibits:\\
(i) monotonic decay of $\tau_c^{F,\gamma}$ and vanishing at
$L_2/\xi_S^0=0.7$ (case $\tilde{\gamma}=0.67$ in Fig.~\ref{fig7}(b));\\
(ii) reentrant behavior characterized by the vanishing of
$\tau_c^{F,\gamma}$ in a certain interval $L_2/\xi_S^0$ ($1<L_2/\xi_S^0<8.2$
for $\tilde{\gamma}=0.75$ in Fig.~\ref{fig7}(b)) and finite $\tau_c^{F,\gamma}$
otherwise;\\
(iii) for the larger values of $\tilde{\gamma}$ ($\tilde{\gamma}=0.9$ in
Fig.~\ref{fig7}(b)), $\tau_c^{F,\gamma}$ first decays to a finite value
exhibiting a minimum at a particular $L_2$ ($L_2/\xi_S^0 \approx 3$), and then
increases again with a further increase of the S-thickness, approaching a
saturation value for thick S-layers. This behavior has a simple physical
explanation: once the domain state is expanded down to $T=0$ (yielding
$\tau_c^{F,\gamma}=0$ at $\tilde{\gamma}=0.75$ and $L_2/\xi_S^0=1.0$) the further
increase of $L_2$ for strong coupling leads to the larger spatial inhomogeneous
terms in the FE-energy (\ref{f_ferr}), which results in a slight enhancement of the
suppressed state towards higher $T$ and as a result, to an increase of
$\tau_c^{F,\gamma}$ (see the two phase diagrams for $L_2/\xi_S^0=5$ and
$L_2/\xi_S^0=15$ shown in Fig.~\ref{fig5}(a)).

\section{CONCLUSIONS\label{conclusions}}

In this paper we have developed a phenomenological model describing periodic
ferroelectric-superconductor multilayers. This model generalizes in some
aspects the microscopic approach proposed in Ref.~\onlinecite{pavlenko} where
the superconducting layers are represented by two planes with a possibility of
charge transfer between the planes and the interior of the layer. We analyze
here the case when the ferroelectric transition temperature lies below the
temperature of the transition to the superconducting state ($T_c^F<T_c^S$) which is
realized in some complex insulating perovskite alloys such as
Ba$_x$Sr$_{1-x}$TiO$_3$ (BST) coupled to high-temperature superconductors.
Similarly to the microscopic model, we observe the stabilization of the
symmetric domain-type phase exhibiting a weak polarization due to the FE-S
contacts above the bulk ferroelectric transition temperature. On the other hand, we
find the polarization-induced suppression of the superconductivity in
the S-layer. We show that the ultimate reason for the destruction of the
superconducting state for strong FE-S coupling, is the increase of the
contribution of the spatial inhomogeneities to the energy of the system.

We study the behavior for different thicknesses of the FE and S-layer
demonstrating the strong change of the topology of the phase diagrams
($\tilde{\gamma}$, $\tau$) with a variation of $L_1/\xi_F^0$ and $L_2/\xi_S^0$.

In general, we find several types of $T_c^{F,\gamma}(L_1)$ and
$T_c^{F,\gamma}(L_2)$ behavior. For the weak FE-S coupling,
$\tilde{\gamma}<\tilde{\gamma}_c$,
we obtain a monotonic variation of $T_c^{F,\gamma}$, with approaching
saturation value in the system with the thick FE or S-layer. In contrast to this, for
the strong coupling, $\tilde{\gamma}>\tilde{\gamma}_c$, the dependence of
$T_c^{F,\gamma}$ vs layer thicknesses appears to be highly nonmonotonic due to the
competition between the first tendency to expand the
domain state (prevailing for the thick S-layers) and the second tendency to
destroy the superconductivity dominating for the thicker FE-layers.

\section*{ACKNOWLEDGEMENTS}

This work has been supported by the Schwerpunktprogramm of the German Science Foundation (Deutsche
Forschungsgemeinschaft) under Grant No.~SPP-1056.

\appendix
\section*{Appendix: Determination of phase $\phi$ in the case without magnetic field}

Let us consider in more details the superconducting layer without a magnetic
field. The general form of the Ginzburg-Landau functional in the S-layer can be
written as
\begin{eqnarray}\label{f_s_g}
&&F_{S}=\int_{L_1}^{L} dz \left[ \frac{1}{2}a_S |\psi(z)|^2+
\frac{1}{4}b_S  |\psi(z)|^4 \right.\\
&&\left. +\frac{\hbar^2}{2m e} \left|\left(\frac{d}{dz}-\frac{2ie}{\hbar c}
A(z)\right)\psi(z)\right|^2 \right],\nonumber
\end{eqnarray}
where $\psi(z)=|\psi(z)| \exp(i\phi(z))$.

Variation of (\ref{f_s_g}) with respect to $A$, $\psi$ and
$\psi^*$ gives the following Ginzburg-Landau equations for the
determination of $A$, $|\psi|$ and $\phi$:
\begin{eqnarray}\label{GL_a}
&&\frac{2e}{\hbar c} A(z)-\phi'(z)=0,\\
&&(a_S+b_S |\psi|^2) |\psi| \cos \phi-\frac{\hbar^2}{m e}\nonumber\\
&&\times(\psi''\cos \phi-\psi' \phi' \sin \phi -(\phi')^2 \psi
\cos \phi)=0, \nonumber \\
&&(a_S+b_S |\psi|^2) |\psi| \sin \phi-\frac{\hbar^2}{m e} \nonumber\\
&&\times(\psi''\sin \phi+\psi' \phi' \cos \phi -(\phi')^2 \psi \sin \phi)=0.
\nonumber
\end{eqnarray}

Note that in the obtained in this case boundary conditions the terms containing
$\phi$ and $\phi'$ appear with the opposite signs and thus cancel, resulting in
the standard form given by (\ref{bz_L1}) and (\ref{bz_L}) in the paper.

From (\ref{GL_a}) we immediately obtain that the single possible solution for
$\phi$ is:
\begin{equation}\label{phi}
\phi=\phi_0=const, \quad L_1<z<L.
\end{equation}
We assume here for simplicity the absence of the tunneling current through the
insulating FE-layer, which corresponds to the zero phase difference on the
contacts between the S-films:
\begin{equation}\label{del_phi}
\Delta \phi=\phi (L_1^+)-\phi (0^-)=0,
\end{equation}
and leads to the constant phase in the whole system: $\phi \equiv \phi_0$. As
the result, we get from (\ref{GL_a})
\begin{equation}\label{A_z}
A(z)=\frac{\hbar c}{2e} \phi'(z)=0,
\end{equation}
and the next two equations transform into the equation
(\ref{g_lan_eq2}) for the absolute value of the superconducting
order parameter $|\psi(z)|$.

\end{document}